# Impact of Radio Frequency Power on Columnar and Filamentary Modes in Atmospheric Pressure Very Low Frequency Plasma within Pores

Haozhe Wang[1], Yu Zhang[1], Jie Cui[1], Zhixin Qian[1], Xiaojiang Huang[1,2,3],

Yu Xu[1,2,3], Jing Zhang[1, 2, 3, *]

1 College of Physics, Donghua University, Shanghai 201620, People's Republic of China

2 Textile Key Laboratory for Advanced Plasma Technology and Application, China National Textile & Apparel Council, Shanghai 201620, People's Republic of China

3 Magnetic Confinement Fusion Research Center, Ministry of Education of the People's Republic of China, Shanghai 201620, People's Republic of China

Email: jingzh@dhu.edu.cn (J.Z.)

## Abstract

The impact of radio frequency (RF) power on columnar and filamentary modes of very low frequency (VLF) plasma within pores is investigated in this work. The 12.5 kHz VLF discharge under various RF powers (13.56 MHz) was analyzed using optical photography and current-voltage measurements. Two-dimensional electron densities were derived using optical emission spectroscopy combined with collisional radiation modeling methods. It is found that RF power and very low frequency voltage ($V_{VLF}$) significantly influence the plasma and its discharge modes within the 200 μm pore. Under low $V_{VLF}$ conditions, the plasma is more intense within the pore, and the discharge mode is columnar discharge. With increasing RF power, the reciprocal motion of electrons counteracts the local enhancement effect of columnar discharge, the discharge transforms into RF discharge, the pore is completely wrapped by the sheath, and the plasma inside is gradually quenched. Under high $V_{VLF}$ conditions, the electron density within the pore is low and the discharge mode is filamentary discharge. RF introduction reduces plasma intensity within the pores firstly. As RF power increases, more ion trapping in the pore increases the field strength distortion and enhances the plasma intensity inside the pore, this enhancement effects becomes more obvious with increasing RF power. In addition, the above effects were observed for all pore widths from 100 μm to 1000 μm. These findings provide key insights for controlling plasma in pores and offer new methodologies for plasma technology applications.

1. Introduction

Porous materials are versatile and play critical roles in areas such as textiles [1], energy [2], environmental protection [3], healthcare [4], and industrial processes [5]. Their ability to interact with gases, liquids, and solids due to their porous structure with huge specific area makes them indispensable across these fields. The external surface and the internal pores of porous materials significantly impact its functionality in storage, transport, and reaction [6]. Atmospheric pressure plasma is known for its high efficiency, environmental friendliness, cleanliness and energy conservation, which finds extensive applications in porous materials like surface modification [7], film deposition [8], new materials synthesis [9] and gas treatment [10] etc. For instance, Zhu et al [11] significantly improved carbon dioxide conversion and energy efficiency by employing a packed-bed dielectric barrier discharge with $Al_2O_3$ and $ZrO_2$ dielectric rods as packing materials. Qin et al [12] deposited a layer of $SiO_xC_yH_z$ film on the top and internal fiber surfaces of a porous polyethylene (PE) separator, which reduced the interfacial resistance of lithium-ion batteries and improved the battery charging and cycling performances. Zhang et al [7] improved the surface properties of polyester fabrics by surface treatment using atmospheric pressure air/helium plasma, which improved the color strength and pigment adhesion of the treated surfaces.

However, the application of atmospheric pressure plasma to porous materials presents a lot challenges, such as the uneven treatment effects in the external surface and the internal pores of porous materials or the damages of pore walls and its overall texture. These problems are mainly related with the short average free path of atmospheric pressure plasma, the large specific surface area of porous materials, the various small internal pores. These factors limit plasma penetration into the pores during atmospheric glow discharge and can damage the porous material when strong filamentary discharges are utilized. In this way, the non-uniform interaction between atmospheric plasma and porous media, leading to uneven internal and external modification effects, has become a key factor restricting the application of atmospheric plasma in porous materials. It is necessary to deeply understand the mechanism and law of interaction between plasma and porous media, and to explore the internal physical mechanism affecting the uniformity of internal and external modification of microporous media, so as to help to solve the problems that have long plagued the application of atmospheric pressure plasma with the porous materials.

Many researchers have recently sought to develop solutions that can effectively address these issues. Armenise et al [13] explored the behavior of plasma inside open-cell polyurethane foams through different arrangement of experimental setups, where functionalized films containing carboxylic acid groups were deposited both inside and outside of the foams, to achieve a uniform coverage of the entire plasma on the substrate. Buddhadasa et al [14] employed a series of alumina-coated copper wires as high-voltage electrodes, which were inserted into the activated carbon in order to generate plasma inside the material, thereby enhancing the selective adsorption capacity of the activated carbon.

There are also many studies on the behavior of plasma in pores, Zhang et al [15, 16] demonstrated by simulating the propagation of plasma in micropores that an increase in the applied voltage significantly enhances the plasma in the pores, and different pore shapes also have an effect on the plasma. It has also been demonstrated that different catalyst materials will have different chemical effects and different dielectric constants, which affect the charge accumulation of the plasma on the dielectric surface and hence the plasma behavior [17, 18].

In addition, some recent studies have highlighted the capabilities of dual-frequency plasma in significantly broadening the operational scope of α-mode discharges and in precisely controlling a wider spectrum of gas temperatures [19-22].The low frequency source generates a transient plasma, forming an electrode sheath and producing a pronounced excitation in the vicinity of the substrate. The radio frequency (RF) source is pivotal in enhancing plasma uniformity by confining electrons within the air gap and fostering pre-ionization, thereby modulating discharge progression [21]. Moreover, experimental and computational analyses have consistently demonstrated that dual-frequency stimulation can effectively modulate electron kinetics and discharge generation in atmospheric-pressure plasmas, optimizing plasma parameters such as electron density and temperature [20-25]. However，the propagation of dual-frequency plasma in pores needs to be investigated further more.

In this study, the impact of single and dual frequency excitation of 12.5 kHz VLF and 13.56 MHz on the plasma inside the pores was investigated. The pores were developed utilizing two alumina ceramic cylinders closely placed. The plasma inside the pores is analyzed through optical photography, voltage and current waveforms analysis of its discharge. Meanwhile, the two-dimensional distribution of the electron density in the pores was diagnosed using the two-dimensional optical emission spectroscopy with a collisional radiative model (2D OES-CRM) method. This study further advances the knowledge in the domain of porous material reaction with the dual-frequency atmospheric dielectric barrier discharges, providing novel perspectives to this field.

## 2. Experiment setup and diagnostic method

### 2.1. Experiment setup

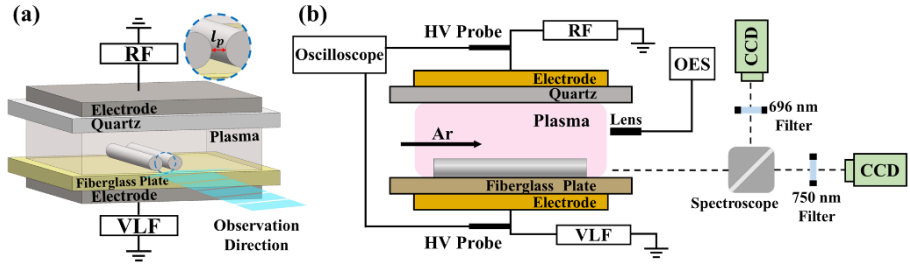

Figure 1. Schematic diagram of the plasma reactor (a) and the 2D OES-CRM experimental device (b)

A schematic diagram of the experimental plasma reactor is presented in figure 1(a). The dielectric barrier discharge (DBD) reactor mainly includes a plate-shaped high-voltage electrode, a plate-shaped RF electrode and two dielectric barrier plates. The radio frequency electrode material is 304 stainless steels, and the size is 50 × 50 mm$^2$. A copper plate of the same size is used as the low frequency high-voltage electrode. The upper dielectric barrier plate material is quartz, the lower dielectric barrier plate material is FR-4 fiberglass and both thickness is 1mm and the gas gap is set to 3 mm. Owing to the superior thermal stability and exceptional chemical inertness inherent to alumina materials, two alumina cylinders with a dielectric constant of 9.8 are situated in the barrier plate gap. The interfiber pores of the fabrics are usually of micrometer scale, ranging from a few micrometers to several hundred micrometers [26]. Therefore, the pore widths $l_p$ were chosen to be 0, 100, 200, 500 and 1000 μm. Each cylinder measures 50 ± 0.5 mm in length, with the end face diameter of 1 ± 0.02 mm. Argon is used as the discharge gas. In this paper, the electron density of the dielectric barrier discharge at atmospheric pressure is close to 10$^{10}$ cm$^{-3}$, and the typical electron temperature at atmospheric pressure is close to 1 eV. According to the Debye length $\lambda_D = \sqrt{\frac{\varepsilon_0 T_e}{e n_e}}$ , the Debye length is calculated to be about 74 μm. Beside the 0 μm pore width, the other pore widths are larger than the Debye length.

A schematic diagram of the 2D OES-CRM experimental device is presented in figure 1(b).

Depending on their frequencies, VLF and RF signals are generated with different systems. The VLF electrode is driven by an AC high-voltage power supply (Suman CTP-2000K) with an output voltage of 0–30 kV and a frequency of 10-15 kHz. The radio frequency electrode is driven by a RF high-voltage power supply (Rishige RSG500S) with an output power of 0-500 W and a frequency of 13.56 MHz. On each electrode, the applied voltages are measured with high voltage probes (Tektronix P6015A), while the current signal is measured with a high-frequency current probe (Pearson 2877). These are connected to a digital oscilloscope RTE1024 (200 MHz).

The optical emission spectroscopy with a collisional radiative model (OES-CRM) diagnostic of the discharge employs two distinct optical arrangements. In the first setup, the plasma optical emission within the pores is monitored using an optical emission spectrometer (Avantes AvaSpec-ULS4096CL-EVO). In the other optical arrangement, the total plasma light is split by a beam splitter, two beams of light were separately passed through narrowband filters with center wavelengths of 696 nm and 750 nm, respectively. Subsequently, these beams are captured by two CCD cameras (Basler aCA 1300-60gm). These cameras are configured in external trigger mode and are simultaneously triggered for capturing images by a signal generator (Siglent SDG 1062X).

2.3. 2D OES-CRM method

In the previous study [27, 28], a two-dimensional OES-CRM method was established to diagnose the electron density and electron temperature of Ar plasma. This method was subsequently employed for the investigation of triple-frequency versus dual-frequency capacitively-coupled Ar plasma in a vacuum, as well as dual-frequency plate-to-plate dielectric barrier discharges at atmospheric pressure. The main reacting particles in the Ar plasma include the electrons, the excited atoms, the ground-state atoms, the atomic ions, the molecular ions as well as the molecular excimers. The major kinetic processes include the electron-impact excitation and de-excitation processes of argon atoms in the ground state (1s), the energy level jump processes of electron as well as atomic collisions, radiative jumps between the various energy levels of argon atoms, collisional dissociation of electrons as well as atoms, collisional ionization of electrons, electronic complex excitation, three-body collisions, penning ionization and diffusion-controlled quenching at wall [29].The pertinent excitation energies and statistical parameters have been previously delineated, and the detailed calculation procedure had been given in the previous work.

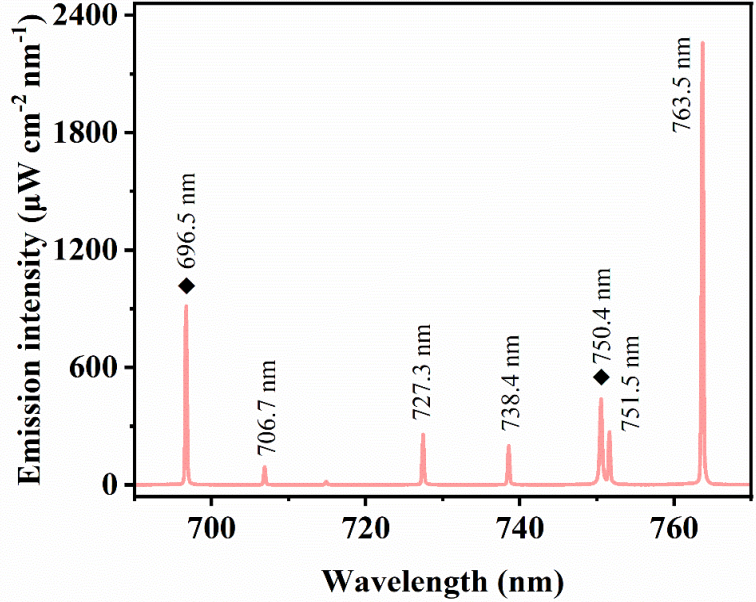

Figure 2. Plasma emission spectra inside the 200 μm pore at $V_{VLF}$ of 11 kV and $P_{RF}$ of 0 W.

Figure 2 shows the plasma emission spectra within a 200 μm pore during atmospheric pressure Ar DBD with $V_{VLF}$ set at 11 kV and $P_{RF}$ (power of radio frequency) at 0 W. Two optical emission lines of the wavelengths of 696.5 nm ($2p_2$ - $1s_5$) and 750.4 nm ($2p_1$ - $1s_2$) were selected to calculate the two-dimensional distribution of electron density within the pore. The parameters of the active species are detailed in table 1. Moreover, due to the inherent instability $V_{VLF}$ atmospheric pressure discharge, two CCD cameras are operated in parallel by a signal generator to capture optical images of the plasma through 696 and 750 ± 5 nm filters. This dual-setup enables the acquisition of two-dimensional distribution of plasma wavelengths at 696 and 750 nm, essentially for accurately determining plasma characteristic. To ensure accuracy, background subtraction at 696 and 750 nm was performed, followed by normalization of the two-dimensional distribution. The absolute radiation intensity of the plasma, measured by OES for procedural calibration and then integrated into the model for computation. The typical duration of plasma treatment of materials is measured in seconds, thus, to prevent errors due to short exposure times, the exposure time for both cameras was set to 0.2 seconds. An aperture of F6.8 was chosen to enhance the depth of field in the photographs, facilitating more comprehensive data acquisition and enhancing the precision of the results. It is worth noting that, despite the proximity of the 750.4 and 751.5 nm lines, the intensity ratio of 750.4 nm to 751.5 nm in atmospheric pressure Ar discharges is approximately 2.1 under various conditions, and the effect of 751.5 nm can be neglected as a result of the normalization that is required after taking the photographs through the filters.

Table 1. Principal emissive species identified for Ar

| λ (nm) | Species | Lower level (eV) | Upper level (eV) | Radiative lifetimes (ns) |
|---|---|---|---|---|
| 696.5 | Ar: $2p_2$-$1s_5$ | 11.548 | 13.328 | 28.3 |
| 706.7 | Ar: $2p_3$-$1s_5$ | 11.548 | 13.302 | 29 |
| 727.3 | Ar: $2p_2$-$1s_4$ | 11.624 | 13.328 | 28.3 |
| 738.4 | Ar: $2p_3$-$1s_4$ | 11.624 | 13.302 | 29 |
| 750.4 | Ar: $2p_1$-$1s_2$ | 11.828 | 13.48 | 21.7 |
| 751.5 | Ar: $2p_5$-$1s_4$ | 11.624 | 13.273 | 24.4 |
| 763.5 | Ar: $2p_6$-$1s_5$ | 11.548 | 13.172 | 29.4 |

## 3. Results and discussion

According to the definition by the International Telecom Union, very low frequency (VLF) is 3 kHz < f < 30 kHz, low frequency (LF) is 30 kHz < f < 300 kHz, middle frequency (MF) is 300 kHz < f < 3 MHz and high frequency (HF) is 3 MHz < f < 30 MHz. According to plasma science, radio frequency (RF) is usually termed for f > 2 MHz. Very low frequency of 10-15 kHz high voltage and RF of 13.56 MHz are applied in this work.

3.1 Plasma behavior of single VLF and RF discharges in the pore

Discharge images are the most intuitive way to observe the characteristics of the discharge in the pores. Figure 3 presents the optical image of a 200 μm pore and pseudo-color discharge images within the pore at varying $V_{VLF}$ of single VLF discharge through a 696 nm bandpass filter. It is observed that discharge within the pore assumes two distinct forms under different voltage conditions. Firstly, when $V_{VLF}$ is low, such as 5 or 6 kV, as shown in figure 3(b) and (c), the plasma is inhomogeneous, an unstable discharge channel is formed between the cylinder and the quartz dielectric plate, due to the low voltage and the electric field distortion caused by the dielectric polarization effect [30]. Meanwhile, the electric field strength of the streamer head is rapidly amplified at the interface between the streamer and the cylinder, resulting in the formation of surface discharges. At the same time, strong but unstable plasma streamers can be observed in the pore. The second case occurs at higher $V_{VLF}$, such as 11 or 14 kV, as shown in figure 3(d) and (e), the plasma is more uniform, a stable and slender discharge channel forms between the cylinder and the quartz dielectric plate, then propagated along the surface of the cylinder. Since the pore width is larger than the Debye length, the plasma is able to propagate into the pore and interact with the pore sidewalls of the cylinder, charging the sidewall surfaces, thus generating an extra electric field along the pore sidewall [17, 31]. Thereby the electric field in the pore is stronger than the electric field outside the pore, and an electric field distortion occurs inside the pore. Here, due to the field strength distortion effect in the pore, which forming a stable plasma streaming light enhancement [15, 16, 18].

It can be observed that two distinct discharge modes exist at varying $V_{VLF}$. At low $V_{VLF}$ in figure 3(b) and (c), the plasma is inhomogeneous within the pore as well as above the alumina cylinder, and a discharge column of about 0.4 mm can be observed at this location, the intensity of the discharge column is stronger than the other regions above the pore. Therefore, it can be considered that the discharge mode under this condition is columnar discharge [32, 33]. It is worth noting that the width of the discharge column is greater than 1 mm for the typical columnar discharge. However, the dielectric polarization effect at the top of the cylinder reduces the width of the discharge column. At high $V_{VLF}$ in figure 3(d) and (e), the plasma is more uniform, but there are still obvious discharge channels at the top of the cylinder. Consequently, the discharge mode can be considered as filamentary discharge [30, 34].

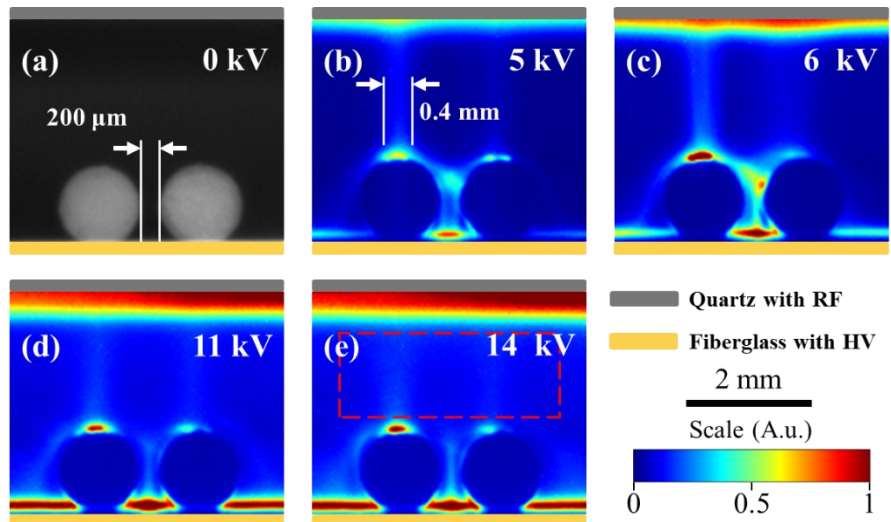

Figure 3. Optical and pseudo-colour images with 200 μm pores through a 696 nm bandpass filter at different $V_{VLF}$ of single VLF discharge

To figure out the discharge mode at different $V_{VLF}$, the voltage-current waveform at different $V_{VLF}$ were plotted in figure 4(a). At low $V_{VLF}$, the current waveform is observed to be smooth, with a minimal number of pulses, and last for a few μs. Concurrently, the plasma within the discharge region was found to be inhomogeneous, enabling the identification of this condition as columnar discharges [32, 35, 36]. At high $V_{VLF}$, discharge pulses lasting less than 1 μs appear on the current waveform, which is characteristics of the typical filamentary discharge [33, 34, 37]. Furthermore, the current signal of the discharge manifests as a series of current pulses. At lower $V_{VLF}$, discharges occur mainly at the voltage rising edge, whereas at higher $V_{VLF}$, dscharges are observed not only at the voltage rising edge but also at its zero-crossing point. Figure 4(b) shows the distribution of the magnitude and the number of current pulses within a single cycle. It is evident that the number of current pulses grows with the increase in $V_{VLF}$, and it can be observed that the magnitude percentage of 0.02 A < I ≤ 0.03 A increases with increasing $V_{VLF}$.

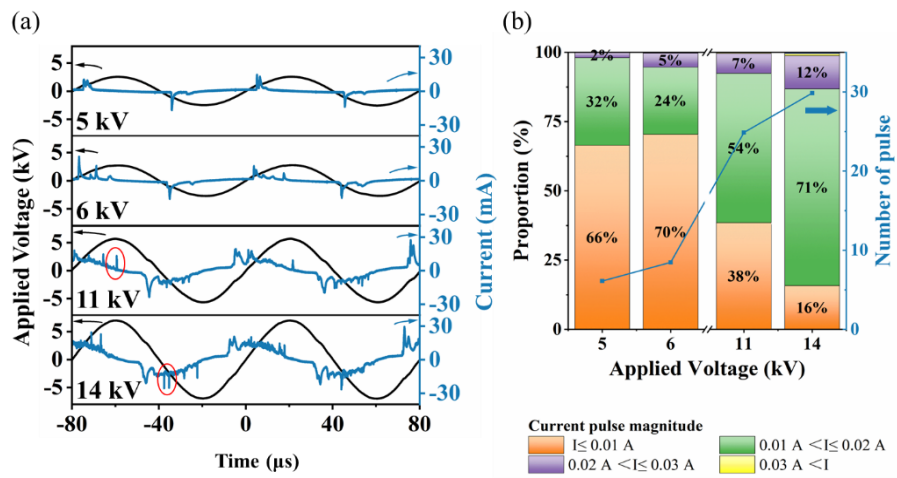

Figure 4. Current characteristics with 200 μm pore at different $V_{VLF}$ of single VLF discharge
(a) Current and voltage waveforms and (b) Current pulse number and magnitude distribution in a single cycle

The pore space between two cylinders were focused for OES-CRM analysis. Figure 5 shows the 2D electron density distribution in 200 μm pore at different $V_{VLF}$ of single VLF discharge. When $V_{VLF}$ is low, the discharges are columnar discharges with inhomogeneous plasma above the cylinder, exhibiting higher electron density within the pore, as shown in figure 5(a) and (b). However, when $V_{VLF}$ is high, discharges transition to filamentary discharges, with lower electron density in the pore but a more uniform plasma above the cylinder, as shown in figure 5(c) and (d). And slightly raising $V_{VLF}$ within the same mode results in a subsequent increase in electron density within the pore. It is also able to observe that the plasma on the bottom of the pore has a higher electron density as it is close to the lower dielectric plate.

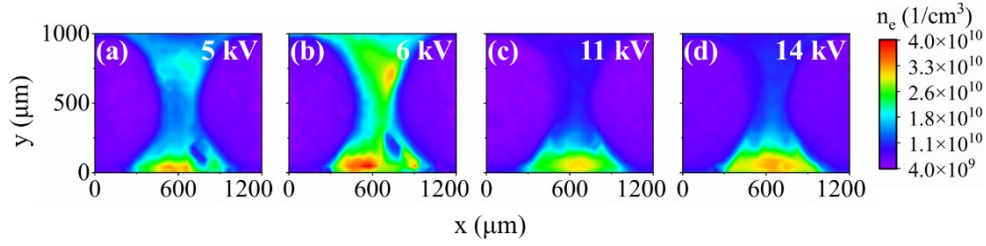

Figure 5. Two-dimensional distribution of electron density with 200 μm pore at different $V_{VLF}$ of single VLF discharge

For other regions, as shown in table 2, the mean, standard deviation, maximum and minimum values of the greyscale in the region above the pores for different $V_{VLF}$ were counted and the selected region is shown in figure (3e). At low $V_{VLF}$, the mean grayscale above the pores is low, which implies that mean intensity of the discharge in the region above the pore is weak, while the high standard deviation of greyscale indicates that the discharge is not uniform. At high $V_{VLF}$, the mean grayscale increases and the discharge is stronger, while the decrease in the standard deviation of greyscale indicates that the plasma is more uniform in the region above the pore.

Table 2. Grayscale analysis of the region above the pore

| $V_{VLF}$ (kV) | Mean (a.u.) | Standard Deviation (a.u.) | Max (a.u.) | Minimum (a.u.) |
| --- | --- | --- | --- | --- |
| 5 | 7.632 | 7.950 | 44 | 0 |
| 6 | 15.661 | 10.580 | 57 | 4 |
| 11 | 27.460 | 5.615 | 49 | 17 |
| 14 | 31.806 | 5.122 | 51 | 19 |

It has been observed that the columnar discharge plasma is inhomogeneous with a higher electron density in the discharge column, which is attributed to the localized enhancement effect. Simulation results show that the development of a column discharge is accompanied by a radial electric field component surrounding the discharge column, which exerts an inhibitory effect on discharge activity in the peripheral region. This radial electric field, directed radially outward from the axis of the discharge column, caused electrons in the surrounding area to move towards the discharge column, thereby impeding discharge development in the adjacent areas [38]. This is consistent with the low mean but high standard deviation of the greyscale at low $V_{VLF}$, as observed in table 2.

With increasing $V_{VLF}$, the decrease of the electron density in the pores in figure (5b) and (c) as well as the decrease in standard deviation of the greyscale above the pores in table 2 indicate that increasing the $V_{VLF}$ leads to a translation in discharge mode from inhomogeneous columnar discharges to diffuse discharges, resulting in a decrease in electron density, which is consistent with the simulation results [36, 39]. Due to the field strength distortion [18] of the pore, the plasma is trapped in the pore, and thus the electron density is higher when the $V_{VLF}$ is lower. As the $V_{VLF}$ increases, the discharge becomes more uniform and dispersed throughout the plasma space, leading to a lower electron density in the pore compared to the inhomogeneous columnar discharges. The electron density increases as the $V_{VLF}$ continues to rise, as shown in figure 5(c) and (d).

For RF discharges, it is difficult to breakdown the 3 mm gap by increasing the $P_{RF}$ at atmospheric pressure. However, it has been demonstrated that the dynamics of the residual charged particles in the discharge gap during the discharge ignition play an important role in the discharge ignition and discharge characteristics, which helps to achieve the stable operation of RF discharge [40]. Therefore, in the experiment, the plasma was first generated by increasing the $V_{VLF}$ for breakdown, followed by increasing the $P_{RF}$, and finally turning off the $V_{VLF}$, thus generating a stable RF DBD at a gap of 3 mm.

As shown in figure 6, the pseudo-color images of the plasma under $P_{RF}$ of 50 W and 60 W through a 696 nm bandpass filter are presented. A distinct sheath with the thickness of about 170 μm can be identified close to the cylinder. In contrast to VLF discharges, RF discharges do not generate plasma within the pore. It was found that the thickness of the sheath at atmospheric pressure RF DBD was about 500 μm (α mode) to 100 μm (γ mode) [41]. In this experiment, the 200 μm pore was completely contained in the sheath, and plasma generation in the pore was suppressed, where plasma tends to be quenched.

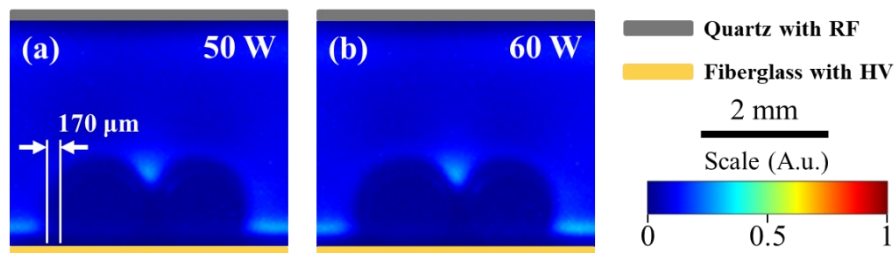

Figure 6. Pseudo-colour images with 200 μm pores through a 696 nm bandpass filter at $P_{RF}$ of 50 W and 60 W of single RF discharge

3.2 Radio frequency impact on columnar and filamentary discharges in the pore

Optical photographs were captured through a 696 nm bandpass filter to delineate the evolution of the plasma distribution in the pore and around the cylinder at different $P_{RF}$. Pseudo-color images of plasma in the 200 μm pore at various $P_{RF}$ from 0 to 60 W when $V_{VLF}$ is set to 6 kV are shown in figure 7. At $P_{RF}$ of 0 W, as shown in figure 7(a), clear discharge columns are observed around the cylinder, and the discharge mode at this point is a column discharge. As shown in figure 7(b) and figure 7(c), with the increase of $P_{RF}$, the plasma is markedly suppressed, and the luminosity within the pore as well as between the top of the cylinder and the upper dielectric plate gradually dims. When the $P_{RF}$ increases to 30 W, as shown in figure 7(d), plasma streamers become indistinguishable, and brightness of the pores approximates that at the top of the cylinder. As shown in figure 7(e), when the $P_{RF}$ is 40 W, the light emission at the edge of the sheath can be slightly observed above the alumina cylinder. Further increasing the $P_{RF}$ to 50 W and 60 W, as shown in figure 7(f) and figure 7(g), the light emission from the edge of the sheath becomes more pronounced. Notably, no light emission is detected within the pore, suggesting that the pore is entirely wrapped by the sheath. Under this condition, the discharge within the pore is suppressed, and the discharge imagery resembles that observed during RF discharges.

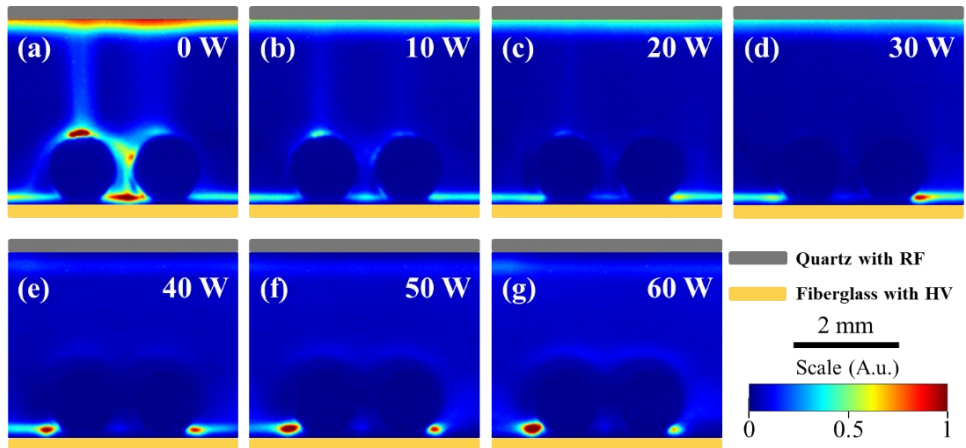

Figure 7. Pseudo-colour images with 200 μm pores through a 696 nm bandpass filter at $V_{VLF}$ of 6 kV and various RF powers

In order to deeply investigate the variation of the plasma electron density in the pore under these conditions, the electron density distribution within a 200 μm pore at a $V_{VLF}$ of 6 kV and $P_{RF}$ ranging from 0 to 60 W are shown in figure 8. It is observed that the electron density within the pore gradually decreases with the increment of $P_{RF}$, thereby suppressing the discharge. Once the $P_{RF}$ reaches 30 W, the electron density inside the pore reaches the lowest value set by the program, $4\times10^9$ cm$^{-3}$, due to the fact that the pore becomes so dark that that the cylinders and the pore are indistinguishable.

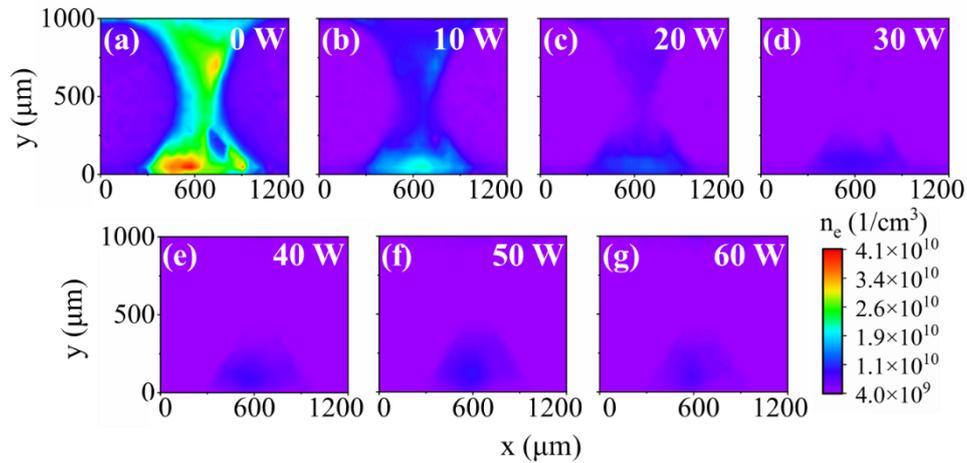

Figure 8. Two-dimensional distribution of electron density within 200 μm pore at $V_{VLF}$ of 6 kV and various RF powers

As $V_{VLF}$ increases to 11 kV, the plasma become more uniform and the discharge mod transforms into a filamentary discharge. Pseudo-color images of plasma in the 200 μm pore at various $P_{RF}$ from 0 to 60 W when $V_{VLF}$ is set to 11 kV are shown in figure 9. When $P_{RF}$ is 0 W, as shown in figure 9(a), plasma streamers are observed around the cylinder and within the pore. When the $P_{RF}$ is increased to 10 W, as shown in figure 9(b), the plasma above the cylinder is rapidly suppressed, while the discharge intensity within the pore is also reduced. Contrary to $V_{VLF}$ of 6 kV, the brightness of the plasma in the pore gradually increases with $P_{RF}$. At $P_{RF}$ of 30 W, as shown in figure 9(d), the brightness in the pore exceeds that at $P_{RF}$ of 0 W. It is also observed that the brightest region near the upper dielectric plate gradually moves downward as the $P_{RF}$ increases, and it can be considered that the introduction of RF limits the movement of ions towards the dielectric surface, which is mainly attributed to the limitation of the mean electric field in the sheath by the introduction of RF [22].

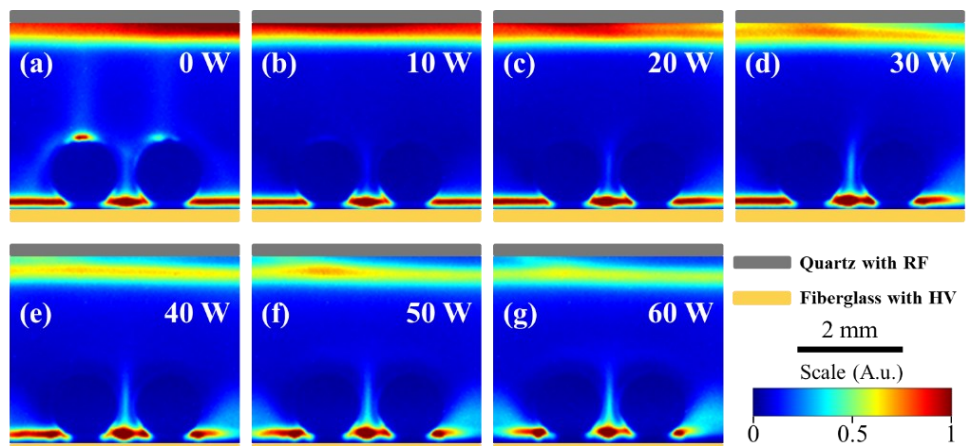

Figure 9. Pseudo-colour images with 200 μm pores through
a 696 nm bandpass filter at $V_{VLF}$ of 11 kV and various RF powers

After analysis by the OES-CRM method, the electron density distribution image of the 200 μm pore with 11 kV $V_{VLF}$ and $P_{RF}$ ranging from 0 to 60 W is presented in figure 10. At $P_{RF}$ of 0 W, as shown in figure 10(a), which corresponds to VLF discharge, the electron density inside the pore is approximately $9.9 \times 10$ cm$^{-3}$. Upon increasing the $P_{RF}$ to 10 W, as shown in figure 10(b), there is a rapid decrease in electron density within the pore. As the power further increases, the electron density within the pore also increases progressively. At 30 W, as shown in figure 10(d), the electron density pore surpasses the value at 0 W, reaching $1.1 \times 10$ cm$^{-3}$. After that, with the continued increase in $P_{RF}$, the electron density in the pore also continues to increase.

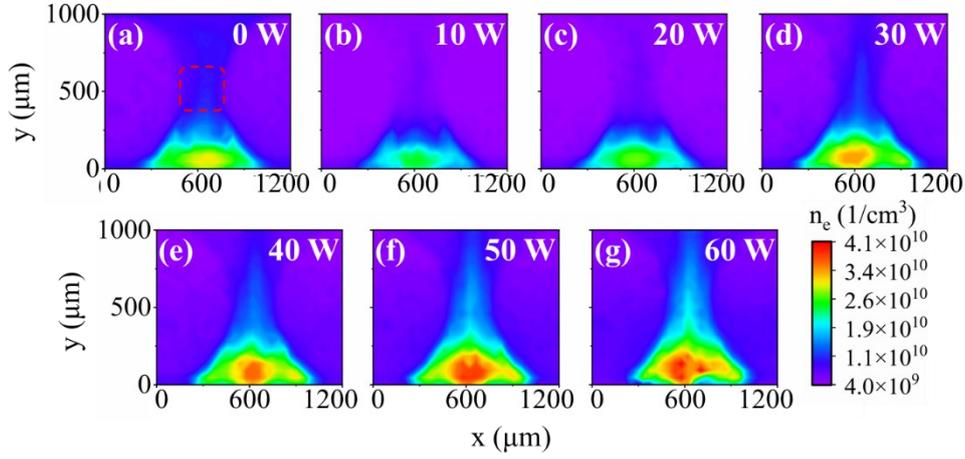

Figure 10. Two-dimensional distribution of electron density
within 200 μm pore at $V_{VLF}$ of 11 kV and various RF powers

In order to visualize the variation of electron density in pores with different $P_{RF}$ at various $V_{VLF}$, the curves of electron density in the 200 μm pore at $V_{VLF}$ of 6 and 11 kV for $P_{RF}$ from 0 to 60 W are illustrated in figure 11. The selected region has been marked in Fig. 10a, which was chosen to be the middle 200 μm of two cylinders. It is evident that the electron density within the pore changes with the increase of $P_{RF}$. At low $V_{VLF}$ of 6 kV, the RF input tends to reduce the electron density within the pore from $2.5 \times 10^{10}$ cm$^{-3}$ at $P_{RF}$ of 0 W to the programmed minimum of $4 \times 10^9$ cm$^{-3}$ at 30 W, at which point the programmed minimum is reached. As the $P_{RF}$ increases to 60 W, the electron density in the pores hardly changes. On the contrary, at high $V_{VLF}$ of 11 kV, the electron density in the pore all shows a tendency to decrease first and then increase with the growth of $P_{RF}$, decreasing from $9.8 \times 10^9$ cm$^{-3}$ at 0 W $P_{RF}$ to $5.5 \times 10^9$ cm$^{-3}$ at 10 W, and then, it gradually grows to $1.70 \times 10^{10}$ cm$^{-3}$ at $P_{RF}$ of 60 W.

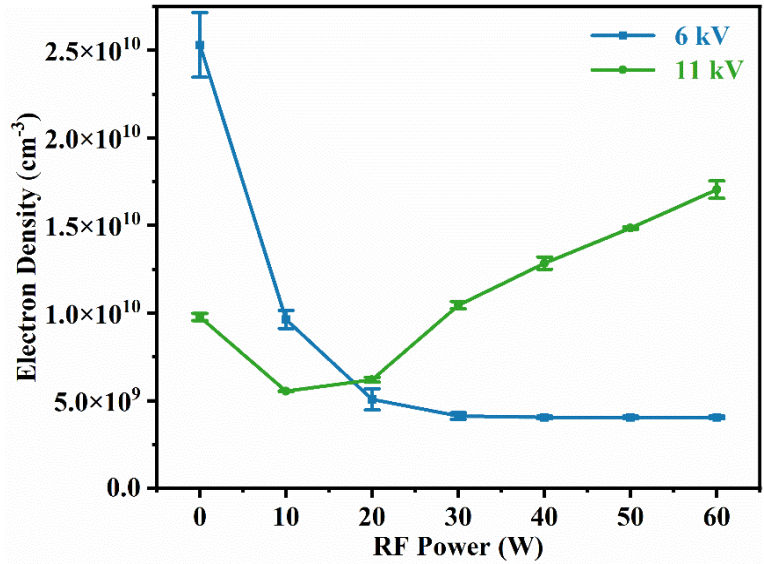

Figure 11. Electron density changes inside 200 um pore at $V_{VLF}$ of 6 and 11 kV with various RF powers

Bazinette et al [24] investigated the impact of applied voltage amplitude and frequency on the plasma in a dual-frequency dielectric barrier discharge. Their findings indicated that when the two frequencies are significantly disparate and the RF voltage is relatively low, the kHz voltage reduces the discharge emission and tends to extinguish the RF discharge. When the $P_{RF}$ is 0 W, the ions in the plasma tend to drift to the transient cathode of the $V_{VLF}$ due to the lower $V_{VLF}$ frequency, the dielectric and pore surfaces are able to accumulate more wall charges, resulting in stronger secondary electron emission [42], which promotes the generation of the next electron avalanche to sustain the plasma.

At low $V_{VLF}$ and when the discharge mode is columnar discharge, the radial electric field induces shrinkage of the discharge area, leading to a stronger discharge at the periphery of the discharge zone. Correspondingly, the discharge accumulated on the dielectric surface at that location is also heightened, which further enhances the discharge in the subsequent voltage half-cycle and encourages the differentiation of the discharge area [38]. It has been demonstrated that the degree of filamentation of the plasma is suppressed when the when RF is introduced [43]. Thus, upon the introduction of 13.56 MHz RF, the electron oscillation frequency increases, the electrons reciprocate with the RF cycle and the electric field changes so fast that the charges in the gap do not have enough time to accumulate on the surface of the dielectric plate. This hindering the accumulation of charges at the dielectric surface, thus inhibiting the local enhancement effect of the discharge and preventing the generation of the discharge column. As the $P_{RF}$ continues to increase, the suppression effect is gradually enhanced. In addition, due to the significant frequency disparity between the VLF and the

RF, the introduction of RF limits the mean electric field in the sheath area and makes the sheath thinner [22], which limits the drift of ions to the VLF transient cathode and reduces the loss of ions from the surface of the dielectric, the ions are trapped by the RF signal within the plasma area [24]. At the same time, the introduction of RF increases the spatially integrated excitation rate and the ionization rate [22], which also helps to increase the ion density in the plasma area. These factors lead to the transition of the discharge from columnar discharges to RF discharges.

As $V_{VLF}$ increases, the discharge mode transforms into filamentary discharges, a higher applied voltage implies a larger oscillation amplitude for electrons and ions. At 10 W RF, the ions still tend to drift towards the transient cathode, particularly when the $V_{VLF}$ is high. However, the RF signal suppresses the drift of ions towards the transient cathode, reducing the number of particles colliding with the transient cathode and consequently decreasing the number of electrons emitted. At the same time, due to the introduction of RF, the electron oscillation frequency increases, and the electrons will be lost more easily on the dielectric surface due to the reciprocal motion between the upper and lower electrode plates at a higher frequency, thus suppressing the memory effect. As a result, the plasma is suppressed and the electron density is reduced.

However, there are simulations confirming that an increase in the applied voltage results in an increase in field strength within the pore, thereby facilitating the ionization of the gas within the pore [15, 18]. Concurrently, the ion density in the plasma area increases as the $P_{RF}$ increases. Since the $V_{VLF}$ is high and the frequency is very low at this time, the ionic oscillation amplitude is still very large and the dielectric surface accumulates a lot of charge generating a stronger secondary electron emission. It has been shown that the strong boundary emission will destroy the RF plasma sheath [44], so the discharge will not be fully transformed into RF mode, being instead dominated by the VLF discharge. Simulations have shown that the direction of the electric field in the VLF discharge is directed towards the pore [16]. The ions will be affected by the electric field and tend to move towards the pore. Therefore, when the $P_{RF}$ is higher than a certain value, the capture of ions in the pore surpasses the loss of electrons to the surface of the dielectric plate. This enhances the electric field distortion, rendering ionization of the gas within the pores more feasible and thereby increasing electron density. However, at the low $V_{VLF}$, the discharge gradually changes to RF mode as the $P_{RF}$ increases, and the resulting sheath wraps around the pore, so the plasma in the pore is suppressed and gradually quenched.

3.3 Pore width effect on dual-frequency plasma in pores

Experimental and simulation studies have demonstrated that the pore width has a significant effect on the discharge within the pore as well as the formation of field strength distortions [15, 16, 45]. It is therefore imperative to explore the impact of RF on filamentary and columnar discharges with various pore widths. The electron densities at 6 or 11 kV $V_{VLF}$, 0, 60 W $P_{RF}$ for pore widths $l_p$ of 0, 100, 200, 500, 1000 μm are presented in figure 12. The introduction of RF results in the suppression of plasma within the pore space at all times when the pore is close to closure. However, for the remaining pore widths, RF suppresses the plasma within the pore at low $V_{VLF}$ and enhances the plasma in the pore at larger $V_{VLF}$. Notably, as the pore width gradually increases from 100 to 1000 μm, the electron density inside the pore decreases under 0 W $P_{RF}$, which corresponds to the conclusion of the existing study [15]. And the reason is that as the pore width increases, the electric field strength decreases in the pore, thus reducing the ionization of the gas in the pore. Furthermore, when the pore width exceeds 500 μm, the electron density in the pore at $V_{VLF}$ of 6 kV, $P_{RF}$ of 60 W is higher compared to that at 200 μm pore. In addition, it has been demonstrated that the field strength distortion is still present in the pores of this case [46], so that the electron density still increases after the RF is introduced. And when the pore width is increased to 1000 μm, there is no significant change in the electron density within the pore under the conditions of $V_{VLF}$ of 6 kV, $P_{RF}$ of 60 W as well as $V_{VLF}$ of 6 kV, $P_{RF}$ of 0 and 60 W as compared to the 500 μm pore. It can be assumed that saturation effects of electron density changes under these conditions have already occurred when the pore width is 1000 μm.

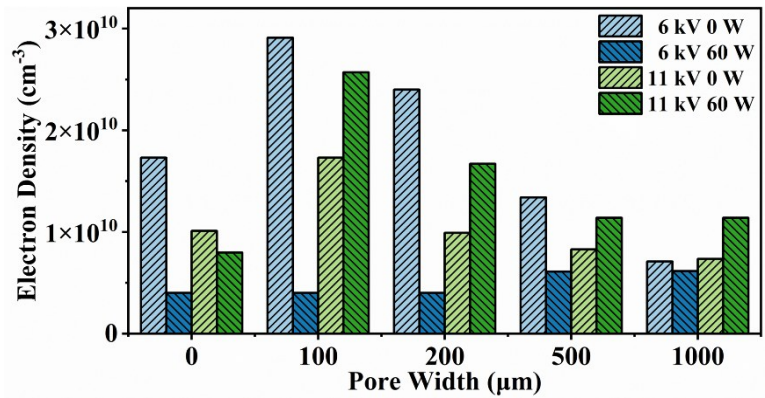

Figure 12. The electron densities at $V_{VLF}$ of 11 and 14 kV at $P_{RF}$ of 0 and 60 W and pore widths of 0, 100, 200, 500, and 1000 μm

## 4. Conclusions

This study demonstrates that RF has significant impact on both columnar and filamentary discharges modes of VLF plasma within pores through electrical diagnosis, optical imaging and two-dimensional OES-CRM methods. The intensity of plasma within 200 μm pore is primarily determined by the very low frequency voltage and the radio frequency power. At low $V_{VLF}$, the plasma within the pore is more intense, and the discharge exhibits an inhomogeneous columnar mode. With introduction of RF and the increase of $P_{RF}$, it is difficult for electrons to accumulate on the dielectric surface, which suppresses the local enhancement effect. The discharge mode is transformed into RF glow, the pore is completely wrapped by the sheath, and the plasma inside the pore is gradually quenched. At high $V_{VLF}$, the current curve displays that the discharge is filamentary. As the introduction of RF and $P_{RF}$ increases, the reciprocating electrons hinder the accumulation of charges on the dielectric surface, and the discharge intensity decreases. However, the discharge mode is dominated by the VLF discharge and more ions tend to move into the pore. When the RF power exceeds a certain value, the discharge inside the pore is enhanced with the electric field inside the pore enhanced. Furthermore, the diminution effects of RF power on the electron density inside the pores during low $V_{VLF}$ columnar discharges and the enhancement effects during high $V_{VLF}$ filamentary discharges are consistently observed with various pore widths of 100, 500 and 1000 μm.

## Data availability statement

All data that support the findings of this study are included within the article (and any supplementary files)

## Acknowledgments

This research was financially supported by the National Natural Science Foundation of China (12075054 and 12205040).

## Authors' contributions

All authors contributed to the study conception and design. Material preparation, data collection and analysis were performed by Haozhe Wang, Yu Zhang, Jie Cui and Zhixin Qian. The first draft of the manuscript was written by Haozhe Wang and all authors commented on previous versions of the manuscript. All authors read and approved the final manuscript.

## Availability of data and material

Not applicable.

Code availability

Not applicable.

Conflict of interest

The authors declare that they do not have any competing for financial interests or personal relationships that could have appeared to influence the work reported in this paper